\documentclass[aps,tightlines]{revtex4}%
\usepackage{amsfonts}
\usepackage{amsmath}
\usepackage{amssymb}
\usepackage{graphicx}%
\providecommand{\U}[1]{\protect\rule{.1in}{.1in}}

\begin{document}
\title{Unitary Invariants and Classification of Four-Qubit States via Negativity Fonts}
\author{S. Shelly Sharma}
\email{shelly@uel.br}
\affiliation{Departamento de F\'{\i}sica, Universidade Estadual de Londrina, Londrina
86051-990, PR Brazil }
\author{N. K. Sharma}
\email{nsharma@uel.br}
\affiliation{Departamento de Matem\'{a}tica, Universidade Estadual de Londrina, Londrina
86051-990 PR, Brazil }

\begin{abstract}
Local unitary invariance and the notion of negativity fonts are used as the
principle tools to construct four qubit invariants of degree 8, 12, and 24. A
degree 8 polynomial invariant that is non-zero on pure four qubit states with
four-body quantum correlations and zero on all other states, is identified.
Classification of four qubit states into seven major classes, using criterion
based on the nature of correlations, is discussed.

\end{abstract}
\maketitle

To detect and quantify entanglement of composite quantum systems is a
challenge taken up with great zeal by theorists and experimentalists alike. On
the way, from the elegant bipartite separability criterion of Peres
\cite{pere96} up to classification schemes for four qubit states
\cite{vers02,vers03,miya03,miya04,ging02,lama07,li09,bors10,vieh11,shar12},
several useful entanglement measures and invariants have been found
\cite{woot98,coff00,dur00,wong01,meye02,vida02,luqu03,oste05,luqu06,shar101,shar102,oste06,heyd04,leva051,leva052,leva06,chte07,doko09,elts12}%
.\ Two qubit entanglement is quantified by\ concurrence \cite{hill97}, which
for a pure state is equal to global negativity \cite{zycz98,vida02}.
Entanglement of a three qubit state due to three-body quantum correlations is
quantified by three tangle \cite{coff00}. For the most general three qubit
state, the difference of squared global negativity and three tangle is a
measure of two qubit correlations and satisfies CKW inequality \cite{coff00}.
A natural question is, which polynomial function of the coefficients
quantifies entanglement due to four-body correlations? Can we write an
invariant analogous to global negativity for two qubits and three tangle for
three qubits to quantify four-body correlations?

Invariant theory describes invariant properties of homogenous polynomials
under general linear transformations. If we write a qubit state in multilinear
form, we can find the set of invariants of the form in terms of state
coefficients $a_{i_{1}i_{2}...i_{N}}$ by using standard methods, as has been
done in \cite{miya03,miya04,luqu03}. One may then investigate the properties
of all invariants in the set. Our general aim, however, is to construct those
polynomial invariants that quantify entanglement due to $K-$body correlations
in an $N-$qubit $\left(  N\geq K\right)  $ pure state. This is done by
constructing $N-$qubit invariants from multivariate forms with $\left(
K-1\right)  -$qubit invariants as coefficients instead of $a_{i_{1}%
i_{2}...i_{N}}$.\ In particular, the invariant that quantifies entanglement
due to $N-$body correlations is obtained from a biform having as coefficients
the $N-1$ qubit invariants. The term $N-$body correlations refers, strictly,
to correlations of the type present in an $N-$qubit GHZ state. The advantage
of our approach \cite{shar101,shar102} is twofold. Firstly, we can choose to
construct invariants that contain information about entanglement of a part of
the system. Secondly, since the form of $N-$qubit invariants is directly
linked to the underlying structure of the composite system state, it can throw
light on the suitability of a given state for a specific information
processing task. Local unitary invariance and the notion of negativity fonts
are used as the principle tools to identify $K-$qubit invariants in an
$N-$qubit state. Negativity fonts are the elementary units of entanglement in
a quantum superposition state. Determinants of negativity fonts are linked to
matrices obtained from state operator through selective partial transposition
\cite{shar07,shar08}. In this article, we obtain analytical expressions for
polynomial invariants of degree $8$, $12$, and $24$ for $N=4$ states. One of
the four qubit invariants is found to be non zero on states with four-body
quantum correlations and zero on separable states as well as on states with
entanglement due to two and three body correlations. It is analogous to three
qubit invariant used to define three tangle \cite{woot98}, and can likewise be
used to construct an entanglement monotone to quantify four-body correlations.

To obtain four qubit invariants that quantify four qubit quantum correlations,
we follow a sequence of steps as given below:

1. Identify two qubit invariants for a given pair in a three qubit state.

2. Obtain a quadratic equation with two qubit invariants for a given pair of
qubits as coefficients. Discriminant of the\ form is the three qubit invariant
written in terms of two qubit invariants.

3. Identify two qubit invariants in a four qubit state. Select three qubits
and write three qubit invariants for these in a four qubit state. We identify
five invariants, including two invariants analogous to ones known for a three
qubit state.

4. A local unitary on fourth qubit yields transformation equations for three
qubit invariants. Proper unitaries can reduce the number of three qubit
invariants in the set to four. The process of finding such local unitaries
yields a quartic equation from which four qubit invariants are obtained. Since
the invariants in a larger Hilbert space are written in terms of relevant
invariants in subspaces, it is possible to differentiate the invariants that
quantify three-body quantum correlations from those that quantify four-body
quantum correlations.

In principle the process can be carried on to higher number of qubits.
Polynomial invariants introduced by Luque and Thibon \cite{luqu03} got
geometrical meaning in the work of Levay \cite{leva06}. We point out the
relation of our four qubit invariants with invariants in \cite{luqu03} and
\cite{leva06}.

Polynomial invariants that identify the nature of correlations in a state are
useful to apply classification criteria proposed in \cite{shar12} to four
qubit states. Two multi qubit pure states are equivalent under stochastic
local operations and classical communication (SLOCC) \cite{dur00} if one can
be obtained from the other with some probability using only local operations
and classical communication amongst different parties. SLOCC equivalence is
the central point in four qubit state classification into nine families in
\cite{vers02}. Borsten et al. \cite{bors10} have invoked the black-hole--qubit
correspondence to derive the classification of four-qubit entanglement.
However, it has been found that the number of four qubit SLOCC\ entanglement
classes is much larger \cite{li09}. The main result of Lamata et al.
\cite{lama07} is that each of the eight genuine in-equivalent entanglement
classes contains a continuous range of strictly non equivalent states,
although with similar structure. O. Viehmann et al. \cite{vieh11} select a set
of generators for the SL(2, C)$\otimes4$ -invariant polynomials or tangles and
classify the eight families of ref. \cite{lama07} using tangle patterns. In
our classification scheme using correlation based criterion \cite{shar12},
multipartite states within the same class have same type of correlations but
may have different number and type of negativity fonts in canonical state (all
the states may not be SLOCC equivalent). In section IV, we calculate the
relevant invariants for SLOCC\ families \cite{vers02}\ and re-classify the
states on the basis of number and nature of negativity fonts with non-zero
determinants. The polynomial invariants used to classify the states in our
scheme quantify correlations generated by distinct interaction types.
Intuitively, this information should be extremely useful to quantum state
engineering. Negativity font analysis can be a helpful tool to optimize the
subsystem interactions to tailor the invariant dynamics for a specific quantum
information processing task. A minor point that will be discussed relates to
the controversy regarding the family L$_{ab_{4}}$ which is pointed out in ref.
\cite{chte07}\ to be a subclass of L$_{abc_{3}}$ with ($a=c)$, while in
\cite{li09} it has been shown that L$_{ab_{4}}$ and L$_{abc_{3}}$ belong to
distinct SLOCC classes.

\section{Negativity Fonts and two qubit invariants}

In this section, we briefly review the concepts of global partial transpose
\cite{pere96}, global negativity \cite{zycz98,vida02}, $K-$way partial
transpose \cite{shar09} and $K-$way negativity fonts \cite{shar101,shar102}.
We also identify those two qubit invariants which determine the entanglement
of a pair of qubits in a three qubit state.

A general N-qubit pure state may be written as%
\begin{equation}
\left\vert \Psi^{A_{1},A_{2},...A_{N}}\right\rangle =\sum_{i_{1}i_{2}...i_{N}%
}a_{i_{1}i_{2}...i_{N}}\left\vert i_{1}i_{2}...i_{N}\right\rangle ,
\label{nstate}%
\end{equation}
where $\left\vert i_{1}i_{2}...i_{N}\right\rangle $ are the basis vectors
spanning $2^{N}$ dimensional Hilbert space, and $A_{p}$ is the location of
qubit $p$. The coefficients $a_{i_{1}i_{2}...i_{N}}$ are complex numbers. The
local basis states of a single qubit are labelled by $i_{m}=0$ and $1,$ where
$m=1,...,N$. The global partial transpose of an $N$ qubit state $\widehat
{\rho}=\left\vert \Psi^{A_{1},A_{2},...A_{N}}\right\rangle \left\langle
\Psi^{A_{1},A_{2},...A_{N}}\right\vert $ with respect to qubit at location $p$
is constructed from the matrix elements of $\widehat{\rho}$ through%
\begin{equation}
\left\langle i_{1}i_{2}...i_{N}\right\vert \widehat{\rho}_{G}^{T_{A_{p}}%
}\left\vert j_{1}j_{2}...j_{N}\right\rangle =\left\langle i_{1}i_{2}%
...i_{p-1}j_{p}i_{p+1}...i_{N}\right\vert \widehat{\rho}\left\vert j_{1}%
j_{2}...j_{p-1}i_{p}j_{p+1}...j_{N}\right\rangle . \label{ptg}%
\end{equation}
To construct a $K-$way partial transpose \cite{shar09}, every matrix element
$\left\langle i_{1}i_{2}...i_{N}\right\vert \widehat{\rho}\left\vert
j_{1}j_{2}...j_{N}\right\rangle $ is labelled by a number $K=\sum
\limits_{m=1}^{N}(1-\delta_{i_{m},j_{m}}),$ where $\delta_{i_{m},j_{m}}=1$ for
$i_{m}=j_{m}$, and $\delta_{i_{m},j_{m}}=0$ for $i_{m}\neq j_{m}$. Matrix
elements of state operator with a given $K$ represent $K-$way coherences
present in the state. Local operations on a quantum superposition transform
$K-$way coherences to $K\pm1$ way coherences. The $K-$way partial transpose
of\ $\widehat{\rho}$ with respect to subsystem $p$ for $K>2$ is obtained by
selective transposition such that%
\begin{align}
\left\langle i_{1}i_{2}...i_{N}\right\vert \widehat{\rho}_{K}^{T_{A_{p}}%
}\left\vert j_{1}j_{2}...j_{N}\right\rangle  &  =\left\langle i_{1}%
i_{2}...i_{p-1}j_{p}i_{p+1}...i_{N}\right\vert \widehat{\rho}\left\vert
j_{1}j_{2}...j_{p-1}i_{p}j_{p+1}...j_{N}\right\rangle ,\nonumber\\
\text{if}\quad\sum\limits_{m=1}^{N}(1-\delta_{i_{m},j_{m}})  &  =K,\quad
\text{and }\quad\delta_{i_{p},j_{p}}=0 \label{ptk1}%
\end{align}
and%
\begin{align}
\left\langle i_{1}i_{2}...i_{N}\right\vert \widehat{\rho}_{K}^{T_{A_{p}}%
}\left\vert j_{1}j_{2}...j_{N}\right\rangle  &  =\left\langle i_{1}%
i_{2}...i_{N}\right\vert \widehat{\rho}\left\vert j_{1}j_{2}...j_{N}%
\right\rangle ,\nonumber\\
\text{if}\quad\sum\limits_{m=1}^{N}(1-\delta_{i_{m},j_{m}})  &  \neq K,
\label{ptk2}%
\end{align}
while%
\begin{align}
\left\langle i_{1}i_{2}...i_{N}\right\vert \widehat{\rho}_{2}^{T_{p}%
}\left\vert j_{1}j_{2}...j_{N}\right\rangle  &  =\left\langle i_{1}%
i_{2}...i_{p-1}j_{p}i_{p+1}...i_{N}\right\vert \widehat{\rho}\left\vert
j_{1}j_{2}...j_{p-1}i_{p}j_{p+1}...j_{N}\right\rangle ,\nonumber\\
\text{if}\quad\sum\limits_{m=1}^{N}(1-\delta_{i_{m},j_{m}})  &  =1\text{ or
}2,\quad\text{and }\quad\delta_{i_{p},j_{p}}=0 \label{pt21}%
\end{align}
and%
\begin{align}
\left\langle i_{1}i_{2}...i_{N}\right\vert \widehat{\rho}_{2}^{T_{p}%
}\left\vert j_{1}j_{2}...j_{N}\right\rangle  &  =\left\langle i_{1}%
i_{2}...i_{N}\right\vert \widehat{\rho}\left\vert j_{1}j_{2}...j_{N}%
\right\rangle ,\nonumber\\
\text{if}\quad\sum\limits_{m=1}^{N}(1-\delta_{i_{m},j_{m}})  &  \neq1\text{ or
}2. \label{pt22}%
\end{align}
One can verify that global partial transpose may be expanded as
\begin{equation}
\widehat{\rho}_{G}^{T_{A_{p}}}=%
{\textstyle\sum\limits_{K=2}^{N}}
\widehat{\rho}_{K}^{T_{A_{p}}}-\left(  N-2\right)  \widehat{\rho}.
\label{decomp}%
\end{equation}
Negativity of $\widehat{\rho}^{T_{p}}$, defined as $N^{A_{p}}=\left(
\left\Vert \rho^{T_{A_{p}}}\right\Vert _{1}-1\right)  ,$where $\left\Vert
\widehat{\rho}\right\Vert _{1}$ is the trace norm of $\widehat{\rho}$, arises
due to all possible negativity fonts present in $\widehat{\rho}^{T_{p}}$.
Since $\widehat{\rho}$ is a positive operators, global negativity depends on
the negativity of $K-$way partially transposed operators with $K\geq2$.

To understand the concept of a negativity font in the context of an $N-$qubit
system, consider the state
\begin{align*}
\left\vert \Phi_{K}^{A_{1}A_{2}...A_{N}}\right\rangle  &  =a_{i_{1}%
i_{2}...i_{N}}\left\vert i_{1}i_{2}...i_{N}\right\rangle +a_{i_{1}%
+1,i_{2}...i_{N}}\left\vert i_{1}+1,i_{2}...i_{N}\right\rangle \\
&  +a_{j_{1}j_{2}...j_{N}}\left\vert j_{1}j_{2}...j_{N}\right\rangle
+a_{j_{1}+1,j_{2}...j_{N}}\left\vert j_{1}+1,j_{2}...j_{N}\right\rangle ,
\end{align*}
with $K=%
{\textstyle\sum\limits_{m=1}^{N}}
\left(  1-\delta_{i_{m}j_{m}}\right)  $ and $\delta_{i_{1}j_{1}}=0$.\ The
state $\left\vert \Phi_{K}^{A_{1}A_{2}...A_{N}}\right\rangle $ is the product
of a $K-$qubit GHZ-like state with $N-K$ qubit product state. Let
$\widehat{\sigma}_{K}^{T_{A_{1}}}$ be the $K-$way partial transpose of
$\widehat{\sigma}_{K}=\left\vert \Phi_{K}^{A_{1}A_{2}...A_{N}}\right\rangle
\left\langle \Phi_{K}^{A_{1}A_{2}...A_{N}}\right\vert $ with respect to qubit
$A_{1}$. If $\widehat{\rho}$ is a pure state given by $\widehat{\rho
}=\left\vert \Psi^{A_{1}A_{2}...A_{N}}\right\rangle \left\langle \Psi
^{A_{1}A_{2}...A_{N}}\right\vert $, then $\widehat{\sigma}_{K}^{T_{A_{1}}}$ is
a $4\times4$ sub-matrix of $\widehat{\rho}_{G}^{T_{A_{1}}}$ and $\widehat
{\rho}_{K}^{T_{A_{1}}}$with negative eigenvalue given by
\[
\lambda^{-}=-\left\vert \det\left[
\begin{array}
[c]{cc}%
a_{i_{1}i_{2}...i_{N}} & a_{j_{1}+1,j_{2}...j_{N}}\\
a_{i_{1}+1,i_{2}...i_{N}} & a_{j_{1}j_{2}...j_{N}}%
\end{array}
\right]  \right\vert .
\]
The matrix $\left[
\begin{array}
[c]{cc}%
a_{i_{1}i_{2}...i_{N}} & a_{j_{1}+1,j_{2}...j_{N}}\\
a_{i_{1}+1,i_{2}...i_{N}} & a_{j_{1}j_{2}...j_{N}}%
\end{array}
\right]  $ is referred to as a $K-$way negativity font \cite{shar101,shar102}.
A symbol used to represent a negativity font, must identify the qubits that
appear in $K$ qubit GHZ-like state. Therefore we split the set of $N$ qubits
with their locations and local basis indices given by, $T=\left\{  \left(
A_{1}\right)  _{i_{1}}\left(  A_{2}\right)  _{i_{2}}...\left(  A_{N}\right)
_{i_{N}}\right\}  ,$ into two subsets, with $S_{1,T}$ containing qubits with
local basis indices satisfying $\delta_{i_{m}j_{m}}=0$ ($i_{m}\neq j_{m}$),
and $S_{2,T}$ having qubits for which $\delta_{i_{m}j_{m}}=1$ ($i_{m}=j_{m}$).
To simplify the notation, we represent by $s_{1,T}$, the sequence of local
basis indices for qubits in $S_{1,T}$. A specific negativity font is therefore
represented by%
\[
\nu_{S_{2,T}}^{i_{1}i_{2}...i_{N}}=\left[
\begin{array}
[c]{cc}%
a_{i_{1}i_{2}...i_{N}} & a_{j_{1}+1,j_{2}...j_{N}}\\
a_{i_{1}+1,i_{2}...i_{N}} & a_{j_{1}j_{2}...j_{N}}%
\end{array}
\right]  .
\]
A nonzero determinant $D_{S_{2,T}}^{s_{1,T}}=\det\left(  \nu_{S_{2,T}}%
^{i_{1}i_{2}...i_{N}}\right)  $ ensures that $\widehat{\sigma}_{K}^{T_{A_{1}}%
}$ is negative. A measurement on the state of a qubit with index in $S_{1,T}$
reduces $\widehat{\sigma}_{K}$ to a separable state, whereas, measuring the
state of a qubit in\ $S_{2,T}$ does not change the negativity of
$\widehat{\sigma}_{K}^{T_{A_{1}}}$. Elementary negativity fonts that quantify
the negativity of $\rho^{T_{A_{p}}}$ for $p\neq1$ are defined in an analogous
fashion. The determinant of a $K-$way negativity font detects $K-$body quantum
correlations in an $N$ qubit state. For even $K$, proper combinations of
determinants of $K-$way negativity fonts are found to be invariant under the
action of local unitary operations on $K$ qubits \cite{shar102}.

For a two qubit state negative eigenvalue of partial transpose is the
invariant that distinguishes between the separable and entangled states.
Global negativity of $\left\vert \Psi^{A_{1}A_{2}}\right\rangle =%
{\textstyle\sum}
a_{i_{1}i_{2}}\left\vert i_{1}i_{2}\right\rangle $ is determined by
$I_{2}^{A_{1}A_{2}}=\left\vert a_{00}a_{11}-a_{01}a_{10}\right\vert $, which
is invariant under $U^{A_{1}}\otimes U^{A_{2}}$. Here $U^{A_{i}}$ is a local
unitary operator that acts on qubit $A_{i}$. The subscript on $I_{2}%
^{A_{1}A_{2}}$ refers to two-body correlations. A two qubit state therefore
has a single negativity font $\nu^{00}=\left[
\begin{array}
[c]{cc}%
a_{00} & a_{01}\\
a_{10} & a_{11}%
\end{array}
\right]  $. In a general three qubit state,
\[
\left\vert \Psi^{A_{1}A_{2}A_{3}}\right\rangle =%
{\textstyle\sum}
a_{i_{1}i_{2}i_{3}}\left\vert i_{1}i_{2}i_{3}\right\rangle ,
\]
the number of two-qubit invariants, for a selected pair of qubits, is three.
For the pair $A_{1}A_{2}$, for example, these are determinants of\ $2-$way
negativity fonts defined as
\begin{equation}
D_{\left(  A_{3}\right)  _{i_{3}}}^{00}=\det\left[
\begin{array}
[c]{cc}%
a_{00i_{3}} & a_{01i_{3}}\\
a_{10i_{3}} & a_{11i_{3}}%
\end{array}
\right]  ,\;i_{3}=0,1,\label{2wayd2}%
\end{equation}
and the difference $\left(  D^{000}-D^{010}\right)  =\left(  D^{000}%
+D^{001}\right)  $, where
\begin{equation}
D^{0i_{2}0}=\det\left[
\begin{array}
[c]{cc}%
a_{0i_{2}0} & a_{0i_{2}+1,1}\\
a_{1i_{2}0} & a_{1,i_{2}+1,1}%
\end{array}
\right]  ,\;i_{2}=0,1,\label{3wayd2}%
\end{equation}
is determinant of a three-way negativity font.

\section{Three-body correlations and three qubit invariants}

Our method was applied in ref. \cite{shar101}, to construct three-tangle
\cite{coff00} and a degree two four qubit invariant which is a function of
determinants of $4-$way negativity fonts. To clarify the process, we review
the three qubit case and show that by using three qubit invariants one may
classify three qubit entangled states into states with i) three and two body
correlations,\ ii) states with only three body correlations and iii) a set of
states with only two body correlations. Class (i) states are the most general
states. Class (ii) states with GHZ\ type entanglement have the form
\[
\left\vert \Psi^{A_{1}A_{2}A_{3}}\right\rangle =a_{i_{1}i_{2}i_{3}}\left\vert
i_{1}i_{2}i_{3}\right\rangle +a_{i_{1}+1,i_{2}+1,i_{3}+1}\left\vert
i_{1}+1,i_{2}+1,i_{3}+1\right\rangle ,
\]
and Class (iii) contains W-like entangled states and bi-separable states of
three qubits. First of all, we write down the transformation equation for two
qubit invariant $D_{\left(  A_{3}\right)  _{1}}^{00}$ to obtain the invariant
which quantifies three-body correlations. The form of this invariant is later
used to identify three qubit invariants in four qubit states. In the absence
of three-body correlations, modified transformation equations yield three
qubit invariants that quantify two body correlations in a three qubit state.

Under a local unitary U$^{A_{3}}=\frac{1}{\sqrt{1+\left\vert x\right\vert
^{2}}}\left[
\begin{array}
[c]{cc}%
1 & -x^{\ast}\\
x & 1
\end{array}
\right]  $, $D_{\left(  A_{3}\right)  _{1}}^{00}$ transforms as%
\begin{equation}
\left(  D_{\left(  A_{3}\right)  _{1}}^{00}\right)  ^{\prime}=\frac
{1}{1+\left\vert x\right\vert ^{2}}\left(  D_{\left(  A_{3}\right)  _{1}}%
^{00}+\left(  x\right)  ^{2}D_{\left(  A_{3}\right)  _{0}}^{00}+x\left(
D^{000}+D^{001}\right)  \right)  ,
\end{equation}
such that%
\begin{equation}
\left(  N_{A_{3}}^{A_{1}A_{2}}\right)  ^{2}=\left\vert D_{\left(
A_{3}\right)  _{1}}^{00}\right\vert ^{2}+\left\vert D_{\left(  A_{3}\right)
_{0}}^{00}\right\vert ^{2}+2\left\vert \left(  \frac{D^{000}+D^{001}}%
{2}\right)  \right\vert ^{2}\label{na1a2}%
\end{equation}
is a three qubit invariant. If the pair of qubits $A_{1}A_{2}$ is entangled
then $N_{A_{3}}^{A_{1}A_{2}}\neq0$. We can verify that global negativity of
$\widehat{\rho}_{G}^{T_{A_{1}}}$ is given by%
\begin{equation}
\left(  N_{G}^{A_{1}}\right)  ^{2}=4\left(  N_{A_{3}}^{A_{1}A_{2}}\right)
^{2}+4\left(  N_{A_{2}}^{A_{1}A_{3}}\right)  ^{2},
\end{equation}
where%
\begin{equation}
\left(  N_{A_{2}}^{A_{1}A_{3}}\right)  ^{2}=\left\vert D_{\left(
A_{2}\right)  _{1}}^{00}\right\vert ^{2}+\left\vert D_{\left(  A_{2}\right)
_{0}}^{00}\right\vert ^{2}+2\left\vert \left(  \frac{D^{000}-D^{001}}%
{2}\right)  \right\vert ^{2}.
\end{equation}
The discriminant of $\left(  D_{\left(  A_{3}\right)  _{1}}^{00}\right)
^{\prime}=0$, yields three qubit invariant
\begin{equation}
I_{3}^{A_{1}A_{2}A_{3}}=\left(  D^{000}+D^{001}\right)  ^{2}-4D_{\left(
A_{3}\right)  _{0}}^{00}D_{\left(  A_{3}\right)  _{1}}^{00}\text{,}%
\label{3way}%
\end{equation}
which is a polynomial invariant of degree four in coefficients $a_{i_{1}%
i_{2}i_{3}}$. The subscript in $I_{3}^{A_{1}A_{2}A_{3}}$ refers to three-body
correlations of the type present in a three qubit GHZ state. The terms
$D^{000}-D^{010}$, $D_{\left(  A_{3}\right)  _{0}}^{00}$, and $D_{\left(
A_{3}\right)  _{1}}^{00}$ vanish on a product state of qubits $A_{1}$ and
$A_{2}$. On the state%
\begin{equation}
\left\vert \Psi^{A_{1}A_{2}}\right\rangle \left\vert \Psi^{A_{3}}\right\rangle
=%
{\textstyle\sum_{i_{1}i_{2}i_{3}}}
a_{i_{1}i_{2}}\left\vert i_{1}i_{2}\right\rangle \left(  b_{0}\left\vert
0\right\rangle +b_{1}\left\vert 1\right\rangle \right)  ;\quad\left(
i_{m}=0,1\right)  ,
\end{equation}
with $D^{00}\neq0$, we have $D_{\left(  A_{3}\right)  _{0}}^{00}=\left(
b_{0}\right)  ^{2}D^{00}$, $D_{\left(  A_{3}\right)  _{1}}^{00}=\left(
b_{1}\right)  ^{2}D^{00}$, and $D^{000}=D^{001}=b_{0}b_{1}D^{00}$ as such
$I_{3}^{A_{1}A_{2}A_{3}}=0.$ Modulus of $I_{3}^{A_{1}A_{2}A_{3}}$, quantifies
the entanglement of qubits $A_{1}A_{2}A_{3}$ due to three body correlations.
Three tangle \cite{coff00}, $\tau_{3}=4\left\vert I_{3}^{A_{1}A_{2}A_{3}%
}\right\vert $, is a well known entanglement monotone.{}

For a general three qubit state with $I_{3}^{A_{1}A_{2}A_{3}}=0$,
\ determinants of two-way fonts transform as%
\[
\left(  D_{\left(  A_{3}\right)  _{0}}^{00}\right)  ^{\prime}=\frac
{1}{1+\left\vert x\right\vert ^{2}}\left(  x^{\ast}\sqrt{D_{\left(
A_{3}\right)  _{1}}^{00}}-\sqrt{D_{\left(  A_{3}\right)  _{0}}^{00}}\right)
^{2},
\]%
\[
\left(  D_{\left(  A_{3}\right)  _{1}}^{00}\right)  ^{\prime}=\frac
{1}{1+\left\vert x\right\vert ^{2}}\left(  x\sqrt{D_{\left(  A_{3}\right)
_{0}}^{00}}+\sqrt{D_{\left(  A_{3}\right)  _{1}}^{00}}\right)  ^{2},
\]
therefore
\begin{equation}
N_{A_{3}}^{A_{1}A_{2}}=\left\vert \left(  D_{\left(  A_{3}\right)  _{0}}%
^{00}\right)  ^{\prime}\right\vert +\left\vert \left(  D_{\left(
A_{3}\right)  _{1}}^{00}\right)  ^{\prime}\right\vert =\left\vert D_{\left(
A_{3}\right)  _{0}}^{00}\right\vert +\left\vert D_{\left(  A_{3}\right)  _{1}%
}^{00}\right\vert ,
\end{equation}
is a three qubit invariant. In other words if $I_{3}^{A_{1}A_{2}A_{3}}=0$ then
$N_{A_{3}}^{A_{1}A_{2}}$ quantifies two body correlations of the pair
$A_{1}A_{2}$. One can verify that $\left\vert D_{\left(  A_{m}\right)  _{0}%
}^{00}\right\vert +\left\vert D_{\left(  A_{m}\right)  _{1}}^{00}\right\vert $
($m=1,2,3$), are three qubit invariants in this case. The sum of product
invariants
\begin{align}
I_{2}^{A_{1}A_{2}A_{3}}  &  =3%
{\textstyle\sum\limits_{\substack{i,j=1\\(i<j)}}^{3}}
\left(  \left\vert D_{\left(  A_{i}\right)  _{0}}^{00}\right\vert +\left\vert
D_{\left(  A_{i}\right)  _{1}}^{00}\right\vert \right)  \left(  \left\vert
D_{\left(  A_{j}\right)  _{0}}^{00}\right\vert +\left\vert D_{\left(
A_{j}\right)  _{1}}^{00}\right\vert \right)  ,\nonumber\\
&  =3\left(  N_{A_{1}}^{A_{2}A_{3}}N_{A_{2}}^{A_{1}A_{3}}+N_{A_{1}}%
^{A_{2}A_{3}}N_{A_{3}}^{A_{1}A_{2}}+N_{A_{2}}^{A_{1}A_{3}}N_{A_{3}}%
^{A_{1}A_{2}}\right)  \label{i23qubit}%
\end{align}
detects W-like tripartite entanglement. It is zero on bi-separable states for
which only one of the three $N_{A_{m}}^{A_{i}A_{j}}=\left\vert D_{\left(
A_{m}\right)  _{0}}^{00}\right\vert +\left\vert D_{\left(  A_{m}\right)  _{1}%
}^{00}\right\vert $ ($i\neq j\neq m$) is non zero and one on a three qubit
W-state. Major classes of three qubits states are uniquely defined by values
of polynomial invariants $4\left\vert I_{3}^{A_{1}A_{2}A_{3}}\right\vert $,
$\left(  N_{G}^{A_{1}}\right)  ^{2}-4\left\vert I_{3}^{A_{1}A_{2}A_{3}%
}\right\vert $, and $I_{2}^{A_{1}A_{2}A_{3}}$.

\section{Four-body correlations and four-qubit invariants}

Four qubit states live in the Hilbert space $C^{2}\otimes C^{2}\otimes
C^{2}\otimes C^{2}$ with a distinct subspace for each set of three qubits. If
there were no four body correlations, three qubit invariants $\left(
I_{3}^{A_{i}A_{j}A_{k}}\right)  _{\left(  A_{l}\right)  _{i_{l}}}(i_{l}=0,1)$,
may determine the entanglement of a four qubit state. In general, additional
three qubit invariants that depend also on four-way negativity fonts exist.
For a selected set of three qubits, three qubit invariants constitute a five
dimensional space and are easily found by the action of a local unitary on the
fourth qubit. To write down transformation equations for three qubit
invariants, first of all, we identify two qubit invariants.

In the most general four qubit state%
\begin{equation}
\left\vert \Psi^{A_{1}A_{2}A_{3}A_{4}}\right\rangle =%
{\textstyle\sum_{i_{1}i_{2}i_{3}i_{4}}}
a_{i_{1}i_{2}i_{3}i_{4}}\left\vert i_{1}i_{2}i_{3}i_{4}\right\rangle
;\quad\left(  i_{m}=0,1\right)  , \label{4qubitstate}%
\end{equation}
when state of qubit $A_{1}$ is transposed, we are looking at entanglement of
qubit $A_{1}$ with rest of the system. Qubit $A_{1}$ may have pairwise
entanglement with qubits $A_{2},$ $A_{3},$ or $A_{4}$. For a given pair, there
are four two-way two qubit invariants (the remaining pair of qubits being in
state $\left\vert 00\right\rangle $, $\left\vert 10\right\rangle $,
$\left\vert 01\right\rangle $ or $\left\vert 11\right\rangle $). For example,
the determinants of two-way negativity fonts for the pair $A_{1}A_{2}$,
written as
\begin{equation}
D_{\left(  A_{3}\right)  _{i_{3}}\left(  A_{4}\right)  _{i_{4}}}^{00}%
=\det\left[
\begin{array}
[c]{cc}%
a_{00i_{3}i_{4}} & a_{01i_{3}i_{4}}\\
a_{10i_{3}i_{4}} & a_{11i_{3}i_{4}}%
\end{array}
\right]  \text{, \ }\left(  i_{3},i_{4}=0,1\right)  ,
\end{equation}
are invariant with respect to unitaries on qubits $A_{1}$ and $A_{2}$.
Three-way coherences generate two qubit invariants $D_{\left(  A_{4}\right)
_{i_{4}}}^{000}-D_{\left(  A_{4}\right)  _{i_{4}}}^{010},\left(
i_{4}=0,1\right)  $, and \ $D_{\left(  A_{3}\right)  _{i_{3}}}^{000}%
-D_{\left(  A_{3}\right)  _{i_{3}}}^{010}\left(  i_{3}=0,1\right)  $, for the
pair $A_{1}A_{2}$. Here determinants of three-way fonts for \{$A_{1}A_{2}%
A_{3}$\} and \{$A_{1}A_{2}A_{4}$\}, respectively, are defined as%
\begin{equation}
D_{\left(  A_{4}\right)  _{i_{4}}}^{0i_{2}0}=\det\left[
\begin{array}
[c]{cc}%
a_{0i_{2}0i_{4}} & a_{0i_{2}+1,1i_{4}}\\
a_{1i_{2}0i_{4}} & a_{1i_{2}+1,1i_{4}}%
\end{array}
\right]  ,\text{ \ }\left(  i_{2},i_{4}=0,1\right)  ,
\end{equation}
and%
\begin{equation}
D_{\left(  A_{3}\right)  _{i_{3}}}^{0i_{2}0}=\det\left[
\begin{array}
[c]{cc}%
a_{0i_{2}i_{3}0} & a_{0i_{2}+1i_{3}1}\\
a_{1i_{2}i_{3}0} & a_{1i_{2}+1i_{3}1}%
\end{array}
\right]  ,\text{ \ }\left(  i_{2},i_{3}=0,1\right)  .
\end{equation}
If four-way negativity fonts are present, then additional $A_{1}A_{2}$
invariants, $D^{0000}-D^{0100}$\ and $D^{0001}-D^{0101}$, are to be
considered. Determinants of four-way negativity fonts are given by%
\begin{equation}
D^{0i_{2}0i_{4}}=\det\left[
\begin{array}
[c]{cc}%
a_{0i_{2}0i_{4}} & a_{0,i_{2}+1,1,i_{4}+1}\\
a_{1i_{2}0i_{4}} & a_{1,i_{2}+1,1,i_{4}+1}%
\end{array}
\right]  ,\text{ \ }\left(  i_{2},i_{4}=0,1\right)  .
\end{equation}
Degree two four qubit invariant%
\begin{equation}
I_{4}=\left(  D^{0000}+D^{0011}-D^{0010}-D^{0001}\right)  , \label{i4}%
\end{equation}
obtained in \cite{shar101} is the same as invariant H of degree two in ref.
\cite{luqu03}. The entanglement monotone, $\tau_{4}=4\left\vert I_{4}%
\right\vert $, was called four-tangle in analogy with three tangle
\cite{coff00}. In \cite{shar102} our method was successfully applied to derive
degree two $N-$qubit invariants for even $N$ and degree four invariants for
odd N in terms of determinants of negativity fonts. It was also shown that one
may use the method to construct $N-$qubit invariants to detect $M-$qubit
correlations ($M\leq N$) in an $N-$qubit state. As an example, we reported
degree four invariants $J_{4}^{\left(  A_{1}A_{2}\right)  }$, $J_{4}^{\left(
A_{1}A_{3}\right)  }$, and $J_{4}^{\left(  A_{1}A_{4}\right)  }$ in ref
\cite{shar102} and found that $\left(  I_{4}\right)  ^{2}=\frac{1}{3}\left(
J_{4}^{\left(  A_{1}A_{2}\right)  }+J_{4}^{\left(  A_{1}A_{3}\right)  }%
+J_{4}^{\left(  A_{1}A_{4}\right)  }\right)  $.

Presently, we focus on the set $A_{1}A_{2}A_{3}$ of three qubits in state
$\left\vert \Psi^{A_{1}A_{2}A_{3}A_{4}}\right\rangle $ (Eq. (\ref{4qubitstate}%
) ) viewed as%
\begin{equation}
\left\vert \Psi^{A_{1}A_{2}A_{3}A_{4}}\right\rangle =\left\vert \Psi_{\left(
A_{4}\right)  _{0}}^{A_{1}A_{2}A_{3}}\right\rangle \left\vert 0\right\rangle
+\left\vert \Psi_{\left(  A_{4}\right)  _{1}}^{A_{1}A_{2}A_{3}}\right\rangle
\left\vert 1\right\rangle ,
\end{equation}
where%
\[
\left\vert \Psi_{\left(  A_{4}\right)  _{i_{4}}}^{A_{1}A_{2}A_{3}%
}\right\rangle =%
{\textstyle\sum_{i_{1}i_{2}i_{3}}}
a_{i_{1}i_{2}i_{3}i_{4}}\left\vert i_{1}i_{2}i_{3}i_{4}\right\rangle
;\quad\left(  i_{4}=0,1\right)  .
\]
Three qubit invariants%
\begin{equation}
\left(  I_{3}^{A_{1}A_{2}A_{3}}\right)  _{\left(  A_{4}\right)  _{i_{4}}%
}=\left(  D_{_{\left(  A_{4}\right)  _{i_{4}}}}^{000}+D_{_{\left(
A_{4}\right)  _{i_{4}}}}^{001}\right)  ^{2}-4D_{\left(  A_{3}\right)
_{0}\left(  A_{4}\right)  _{i_{4}}}^{00}D_{\left(  A_{3}\right)  _{1}\left(
A_{4}\right)  _{i_{4}}}^{00};\quad i_{4}=0,1,
\end{equation}
quantify GHZ\ state like three-way correlations in three qubit subspace
$C^{2}\otimes C^{2}\otimes C^{2}$. Continuing the search for a four qubit
invariant that detects four qubit correlations, we examine the transformation
of three qubit invariant $\left(  I_{3}^{A_{1}A_{2}A_{3}}\right)  _{\left(
A_{4}\right)  _{1}}$under U$^{A_{4}}=\frac{1}{\sqrt{1+\left\vert y\right\vert
^{2}}}\left[
\begin{array}
[c]{cc}%
1 & -y^{\ast}\\
y & 1
\end{array}
\right]  $. The resulting transformation equation is%
\begin{align}
&  \left(  I_{3}^{A_{1}A_{2}A_{3}}\right)  _{\left(  A_{4}\right)  _{1}%
}^{\prime}=\frac{1}{\left(  1+\left\vert y\right\vert ^{2}\right)  ^{2}%
}\left[  y^{4}\left(  I_{3}^{A_{1}A_{2}A_{3}}\right)  _{\left(  A_{4}\right)
_{0}}+4y^{3}P_{\left(  A_{4}\right)  _{0}}^{A_{1}A_{2}A_{3}}\right.
\nonumber\\
&  \left.  +6y^{2}T_{A_{4}}^{A_{1}A_{2}A_{3}}+4yP_{\left(  A_{4}\right)  _{1}%
}^{A_{1}A_{2}A_{3}}+\left(  I_{3}^{A_{1}A_{2}A_{3}}\right)  _{\left(
A_{4}\right)  _{1}}\right]  ,\label{i3}%
\end{align}
where%
\begin{align}
T_{A_{4}}^{A_{1}A_{2}A_{3}} &  =\frac{1}{6}\left(  D^{0000}+D^{0011}%
+D^{0010}+D^{0001}\right)  ^{2}\nonumber\\
&  -\frac{2}{3}\left(  D_{\left(  A_{3}\right)  _{0}}^{000}+D_{\left(
A_{3}\right)  _{0}}^{001}\right)  \left(  D_{\left(  A_{3}\right)  _{1}}%
^{000}+D_{\left(  A_{3}\right)  _{1}}^{001}\right)  \nonumber\\
&  +\frac{1}{3}\left(  D_{_{\left(  A_{4}\right)  _{0}}}^{000}+D_{_{\left(
A_{4}\right)  _{0}}}^{001}\right)  \left(  D_{_{\left(  A_{4}\right)  _{1}}%
}^{000}+D_{_{\left(  A_{4}\right)  _{1}}}^{001}\right)  \nonumber\\
&  -\frac{2}{3}\left(  D_{\left(  A_{3}\right)  _{0}\left(  A_{4}\right)
_{0}}^{00}D_{\left(  A_{3}\right)  _{1}\left(  A_{4}\right)  _{1}}%
^{00}+D_{\left(  A_{3}\right)  _{0}\left(  A_{4}\right)  _{1}}^{00}D_{\left(
A_{3}\right)  _{1}\left(  A_{4}\right)  _{0}}^{00}\right)  ,
\end{align}
and%
\begin{align}
P_{\left(  A_{4}\right)  _{i_{4}}}^{A_{1}A_{2}A_{3}} &  =\frac{1}{2}\left(
D_{_{\left(  A_{4}\right)  _{i_{4}}}}^{000}+D_{_{\left(  A_{4}\right)
_{i_{4}}}}^{001}\right)  \left(  D^{0000}+D^{0011}+D^{0010}+D^{0001}\right)
\nonumber\\
&  -\left(  D_{\left(  A_{3}\right)  _{1}\left(  A_{4}\right)  _{i_{4}}}%
^{00}\left(  D_{\left(  A_{3}\right)  _{0}}^{000}+D_{\left(  A_{3}\right)
_{0}}^{001}\right)  +D_{\left(  A_{3}\right)  _{0}\left(  A_{4}\right)
_{i_{4}}}^{00}\left(  D_{\left(  A_{3}\right)  _{1}}^{000}+D_{\left(
A_{3}\right)  _{1}}^{001}\right)  \right)  .
\end{align}
Discriminant of a quartic equation, $y^{4}a-4by^{3}+6y^{2}c-4dy+f=0$, in
variable $y$ is $\Delta=S^{3}-27T^{2}$ where $S=3c^{2}-4bd+af$, and
$T=acf-ad^{2}-b^{2}f+2bcd-c^{3}$ (cubic invariant ), are polynomial
invariants. When a selected U$^{A_{4}}$ results in $\left(  \left(
I_{3}^{A_{1}A_{2}A_{3}}\right)  _{\left(  A_{4}\right)  _{1}}\right)
^{\prime}=0$ (Eq. (\ref{i3})), the associated polynomial invariant is%
\begin{align}
I_{(4,8)}^{A_{1}A_{2}A_{3}A_{4}} &  =3\left(  T_{A_{4}}^{A_{1}A_{2}A_{3}%
}\right)  ^{2}-4P_{\left(  A_{4}\right)  _{0}}^{A_{1}A_{2}A_{3}}P_{\left(
A_{4}\right)  _{1}}^{A_{1}A_{2}A_{3}}\nonumber\\
&  +\left(  I_{3}^{A_{1}A_{2}A_{3}}\right)  _{\left(  A_{4}\right)  _{0}%
}\left(  I_{3}^{A_{1}A_{2}A_{3}}\right)  _{\left(  A_{4}\right)  _{1}%
},\label{i48}%
\end{align}
which is a four qubit invariant of degree eight expressed in terms of three
qubit invariants for $A_{1}A_{2}A_{3}$. In order to distinguish between degree
$2$ invariant $I_{4}$ and the new invariant, degree of the invariant has been
added to the subscript. By construction, the four qubit invariant
$I_{(4,8)}^{A_{1}A_{2}A_{3}A_{4}}$ is a combination of three qubit
($A_{1}A_{2}A_{3}$) invariants. It is easily verified that on a state which is
a product of $\left\vert \Psi^{A_{1}A_{2}A_{3}}\right\rangle =%
{\textstyle\sum_{i_{1}i_{2}i_{3}}}
a_{i_{1}i_{2}i_{3}}\left\vert i_{1}i_{2}i_{3}\right\rangle $ with
$I_{3}^{A_{1}A_{2}A_{3}}\neq0,$ and $\Psi^{A_{4}}=d_{0}\left\vert
0\right\rangle +d_{1}\left\vert 1\right\rangle ,$ we obtain%

\begin{equation}
\left(  I_{3}^{A_{1}A_{2}A_{3}}\right)  _{\left(  A_{4}\right)  _{0}}=\left(
I_{3}^{A_{1}A_{2}A_{3}}\right)  _{\left(  A_{4}\right)  _{1}}=T_{A_{4}}%
^{A_{1}A_{2}A_{3}}=P_{\left(  A_{4}\right)  _{0}}^{A_{1}A_{2}A_{3}}=P_{\left(
A_{4}\right)  _{1}}^{A_{1}A_{2}A_{3}},
\end{equation}
leading to $I_{(4,8)}^{A_{1}A_{2}A_{3}A_{4}}=0$. Likewise, $I_{(4,8)}%
^{A_{1}A_{2}A_{3}A_{4}}$ vanishes on product state $\left\vert \Psi
^{A_{1}A_{2}}\right\rangle \left\vert \Psi^{A_{3}A_{4}}\right\rangle $, where
$\left\vert \Psi^{A_{1}A_{2}}\right\rangle =%
{\textstyle\sum_{i_{1}i_{2}}}
a_{i_{1}i_{2}}\left\vert i_{1}i_{2}\right\rangle $ and $\left\vert \Psi
^{A_{3}A_{4}}\right\rangle =%
{\textstyle\sum_{i_{3}i_{4}}}
b_{i_{3}i_{4}}\left\vert i_{3}i_{4}\right\rangle $. Besides that
$I_{(4,8)}^{A_{1}A_{2}A_{3}A_{4}}=0$ on a four qubit W-like state, and all
entangled states with only three and two-body correlations, as seen in section IV.

The cubic invariant associated with Eq. (\ref{i3}) is%
\begin{equation}
J^{A_{1}A_{2}A_{3}A_{4}}=\det\left[
\begin{array}
[c]{ccc}%
\left(  I_{3}^{A_{1}A_{2}A_{3}}\right)  _{\left(  A_{4}\right)  _{1}} &
P_{\left(  A_{4}\right)  _{1}}^{A_{1}A_{2}A_{3}} & T_{A_{4}}^{A_{1}A_{2}A_{3}%
}\\
P_{\left(  A_{4}\right)  _{1}}^{A_{1}A_{2}A_{3}} & T_{A_{4}}^{A_{1}A_{2}A_{3}}
& P_{\left(  A_{4}\right)  _{0}}^{A_{1}A_{2}A_{3}}\\
T_{A_{4}}^{A_{1}A_{2}A_{3}} & P_{\left(  A_{4}\right)  _{0}}^{A_{1}A_{2}A_{3}}
& \left(  I_{3}^{A_{1}A_{2}A_{3}}\right)  _{\left(  A_{4}\right)  _{0}}%
\end{array}
\right]  ,
\end{equation}
while the discriminant reads as%
\begin{equation}
\Delta=\left(  I_{\left(  4,8\right)  }^{A_{1}A_{2}A_{3}A_{4}}\right)
^{3}-27\left(  J^{A_{1}A_{2}A_{3}A_{4}}\right)  ^{2}\text{.}%
\end{equation}
Since there are four ways in which a given set of three qubits may be
selected, $\Delta$ can be expressed in terms of different sets of three qubit
invariants. In addition (Eq. (\ref{i3})) also leads to%
\begin{align}
\left(  N_{A_{4}}^{A_{1}A_{2}A_{3}}\right)  ^{2}  &  =\left\vert \left(
I_{3}^{A_{1}A_{2}A_{3}}\right)  _{\left(  A_{4}\right)  _{0}}\right\vert
^{2}+\left\vert \left(  I_{3}^{A_{1}A_{2}A_{3}}\right)  _{\left(
A_{4}\right)  _{1}}\right\vert ^{2}\nonumber\\
+6\left\vert T_{A_{4}}^{A_{1}A_{2}A_{3}}\right\vert ^{2}  &  +4\left\vert
P_{\left(  A_{4}\right)  _{0}}^{A_{1}A_{2}A_{3}}\right\vert ^{2}+4\left\vert
P_{\left(  A_{4}\right)  _{1}}^{A_{1}A_{2}A_{3}}\right\vert ^{2},
\end{align}
which is a four qubit invariant analogous to $\left(  N_{A_{3}}^{A_{1}A_{2}%
}\right)  ^{2}$ (Eq. (\ref{na1a2})) for three qubit states. In general, one
can construct an invariant $N_{A_{l}}^{A_{i}A_{j}A_{k}}$ ($i\neq j\neq k\neq
l$) for a selected three qubit subsystem $A_{i}A_{j}A_{k}$ of four qubit
state. In analogy with global negativity, one may define a four qubit
invariant of degree four,
\begin{equation}
\left(  N_{(4,4)}^{A_{1}}\right)  ^{2}=16\left(  N_{A_{4}}^{A_{1}A_{2}A_{3}%
}\right)  ^{2}+16\left(  N_{A_{3}}^{A_{1}A_{2}A_{4}}\right)  ^{2}+16\left(
N_{A_{2}}^{A_{1}A_{3}A_{4}}\right)  ^{2}, \label{na1_48}%
\end{equation}
which detects bipartite entanglement of qubit $A_{1}$ with subsystem
$A_{2}A_{3}A_{4}$ due to three and four body quantum correlations. If
$I_{(4,8)}^{A_{1}A_{2}A_{3}A_{4}}=0$, but at least two of the $N_{A_{l}%
}^{A_{i}A_{j}A_{k}}$ are finite, then 4-partite entanglement can be due to
three and two body correlations. In this case the invariant that detects
entanglement may be defined as%
\begin{align}
N_{(4,8)}^{A_{1}A_{2}A_{3}A_{4}}  &  =16N_{A_{1}}^{A_{2}A_{3}A_{4}}N_{A_{2}%
}^{A_{1}A_{3}A_{4}}+16\left(  N_{A_{1}}^{A_{2}A_{3}A_{4}}+N_{A_{2}}%
^{A_{1}A_{3}A_{4}}\right)  N_{A_{3}}^{A_{1}A_{2}A_{4}}\nonumber\\
&  +16\left(  N_{A_{1}}^{A_{2}A_{3}A_{4}}+N_{A_{2}}^{A_{1}A_{3}A_{4}}%
+N_{A_{3}}^{A_{1}A_{2}A_{4}}\right)  N_{A_{4}}^{A_{1}A_{2}A_{3}}.
\end{align}
On the other hand, if we have a state on which all $N_{A_{l}}^{A_{i}A_{j}%
A_{k}}$ are zero, then the quantities $I_{A_{r}A_{s}}^{A_{p}A_{q}}=\sum
_{i_{r}i_{s}}\left\vert D_{\left(  A_{r}\right)  _{i_{r}}\left(  A_{s}\right)
_{i_{s}}}^{00}\right\vert $ ($p\neq q\neq r\neq s=1$ to $4$), turn out to be
four qubit invariants. A different class of entangled states is obtained if
only one of the $N_{A_{l}}^{A_{i}A_{j}A_{k}}$ is non zero along with a finite
$I_{A_{r}A_{s}}^{A_{p}A_{l}}$. In section II we noted that $I_{2}^{A_{1}%
A_{2}A_{3}}$ (Eq. (\ref{i23qubit})) detects W like entanglement of three
qubits $A_{1}A_{2}A_{3}$. Likewise, when $I_{(4,8)}^{A_{1}A_{2}A_{3}A_{4}%
}=N_{(4,8)}^{A_{1}A_{2}A_{3}A_{4}}=0$, the invariant%
\begin{align}
I_{(2,6)}^{A_{1}A_{2}A_{3}A_{4}}  &  =\frac{3}{2}\left(  I_{2}^{A_{1}%
A_{2}A_{3}}\right)  \left(  I_{A_{2}A_{3}}^{A_{1}A_{4}}+I_{A_{1}A_{3}}%
^{A_{2}A_{4}}+I_{A_{1}A_{2}}^{A_{3}A_{4}}\right) \nonumber\\
&  +\frac{3}{2}\left(  I_{2}^{A_{1}A_{2}A_{4}}\right)  \left(  I_{A_{1}A_{4}%
}^{A_{2}A_{3}}+I_{A_{1}A_{3}}^{A_{3}A_{4}}\right)  +\frac{3}{2}\left(
I_{2}^{A_{1}A_{3}A_{4}}\right)  I_{A_{1}A_{3}}^{A_{2}A_{4}} \label{i26}%
\end{align}
detects W-like four qubit entanglement. Here $\left(  I_{2}^{A_{p}A_{q}A_{r}%
}\right)  _{A_{s}}=3I_{A_{r}A_{s}}^{A_{p}A_{q}}I_{A_{q}A_{s}}^{A_{p}A_{r}}$,
$(p\neq q\neq r\neq s=1$ to $4),$ is the invariant that detects W-like
entanglement of qubits $A_{p}A_{q}A_{r}$ in a four qubit state.

In ref. \cite{leva06} four qubit invariants have been obtained in terms of
coefficients having geometrical significance. A comparison of Eq. (56) of ref.
\cite{leva06} with our Eq. (\ref{i3}), indicates that their set of invariants
($I_{1}$, $I_{2}$, $I_{3}$, $I_{4}$) may be expressed in terms of our three
qubit invariants, though they are not exactly the same. A method equivalent to
method of Schlafli \cite{gelf94} has been used to arrive at Eq. (22)%
\[
R(\text{t})=c_{0}t_{0}^{4}+4c_{1}t_{0}^{3}t_{1}+6c_{2}t_{0}^{2}t_{1}%
^{2}+4c_{3}t_{0}t_{1}^{2}+c_{4}t_{1}^{4}%
\]
by Luque and Thibon \cite{luqu03}. Then higher degree invariants are expressed
in terms of $c_{i}$ coefficients and computer algebra relates these to basic
four qubit invariants. Since for $t_{0}=1$, expression for $R($t$)$ has the
same form as Eq. (\ref{i3}), a direct correspondence can be established
between $c_{i}$ coefficients and our three qubit invariants. Such a comparison
establishes a neat connection of our invariants with projective geometry
approach and classical invariant theory concepts.

\section{Invariants and Classification of four-qubit States}

Decomposition of global partial transpose $\widehat{\rho}_{G}^{T_{A_{p}}}$ of
four qubit state $\left\vert \Psi^{A_{1}A_{2}A_{3},A_{4}}\right\rangle $ with
respect to qubit $A_{p}$ in terms of $K-$way partially transposed operators
(Eq. (\ref{decomp})) reads as%
\begin{equation}
\widehat{\rho}_{G}^{T_{A_{p}}}=\sum\limits_{K=2}^{4}\widehat{\rho}%
_{K}^{T_{A_{p}}}-2\widehat{\rho}.\label{3n}%
\end{equation}
When a state has only $K-$way coherences, we have $\widehat{\rho}%
_{G}^{T_{A_{p}}}=\widehat{\rho}_{K}^{T_{A_{p}}}$, for a selected set of $K$
qubits. For a given qubit, the number of $K-$way negativity fonts in a $K-$way
partially transposed matrix varies from $0$ to $4$. Local unitary operations
can be used to annihilate the negativity fonts that is obtain a state for
which determinants of selected negativity fonts are zero. The process leads to
canonical state which is a state written in terms of minimum number of local
basis product states \cite{acin00}. In ref. \cite{shar12}, we proposed a
classification scheme in which an entanglement class is characterized by the
minimal set of $K-$way $\left(  2\leq K\leq4\right)  $\ partially transposed
matrices present in the expansion of global partial transpose of the canonical
state. Seven possible ways in which the global partial transpose (GPT) of a
four qubit canonical state may be decomposed correspond to seven major
entanglement classes that is class I. $\left(  \widehat{\rho}_{c}\right)
_{G}^{T_{A_{p}}}=\sum\limits_{K=2}^{4}\left(  \widehat{\rho}_{c}\right)
_{K}^{T_{A_{p}}}-2\widehat{\rho}_{c}$, II. $\left(  \widehat{\rho}_{c}\right)
_{G}^{T_{A_{p}}}=\left(  \widehat{\rho}_{c}\right)  _{4}^{T_{A_{p}}}+\left(
\widehat{\rho}_{c}\right)  _{3}^{T_{A_{p}}}-\widehat{\rho}_{c}$, III. $\left(
\widehat{\rho}_{c}\right)  _{G}^{T_{A_{p}}}=\left(  \widehat{\rho}_{c}\right)
_{4}^{T_{A_{p}}}+\left(  \widehat{\rho}_{c}\right)  _{2}^{T_{A_{p}}}%
-\widehat{\rho}_{c}$ , IV. $\ \left(  \widehat{\rho}_{c}\right)
_{G}^{T_{A_{p}}}=\left(  \widehat{\rho}_{c}\right)  _{4}^{T_{A_{p}}}$, V.
$\left(  \widehat{\rho}_{c}\right)  _{G}^{T_{A_{p}}}=\left(  \widehat{\rho
}_{c}\right)  _{3}^{T_{A_{p}}}+\left(  \widehat{\rho}_{c}\right)
_{2}^{T_{A_{p}}}-\widehat{\rho}_{c}$ , VI. $\left(  \widehat{\rho}_{c}\right)
_{G}^{T_{A_{p}}}=\left(  \widehat{\rho}_{c}\right)  _{3}^{T_{A_{p}}}$, and
VII. $\left(  \widehat{\rho}_{c}\right)  _{G}^{T_{A_{p}}}=\left(
\widehat{\rho}_{c}\right)  _{2}^{T_{A_{p}}}$. Of these, six classes contain
states with four-partite entanglement, while class VI with $\widehat{\rho}%
_{G}^{T_{A_{p}}}=\left(  \widehat{\rho}_{3}^{T_{A_{p}}}\right)  _{c}$ has only
three qubit entanglement. Each major class contains sub-classes depending on
the number and type of negativity fonts in global partial transpose of the
canonical state. Table \ref{t1} lists the decomposition of $\left(  \rho
_{c}\right)  _{G}^{T_{A_{p}}}$, invariants $I_{\left(  4,8\right)  }%
^{A_{1}A_{2}A_{3}A_{4}}$, $D_{A_{4}}^{A_{1}A_{2}A_{3}}$, $\Delta$, and
$N_{K-\text{way}}$ ($K=$2,3,4) in canonical state, for different classes of
four qubit entangled states. Here $D_{A_{4}}^{A_{1}A_{2}A_{3}}=\left(
N_{A_{4}}^{A_{1}A_{2}A_{3}}\right)  ^{2}-2\left\vert I_{(4,8)}^{A_{1}%
A_{2}A_{3}A_{4}}\right\vert $, is a measure of residual three-way correlations
between qubits $A_{1}A_{2}A_{3}$ and $N_{K-\text{way}}$ ($K=$2,3,4) is the
number of $K-$way negativity fonts in a state. \begin{table}[ptb]
\caption{Decomposition of $\left(  \rho_{c}\right)  _{G}^{T_{A_{p}}}$,
invariants $I_{\left(  4,8\right)  }^{A_{1}A_{2}A_{3}A_{4}}$, $D_{A_{4}%
}^{A_{1}A_{2}A_{3}}$, $\Delta$, and $N_{K-\text{way}}$ ($K=$2,3,4) in
canonical state, for seven classes of four qubit entangled states}%
\begin{tabular}
[c]{||c||c||c||c||c||c||c||c||}\hline\hline
Class & Decomposition of $\left(  \rho_{c}\right)  _{G}^{T_{A_{p}}}$ &
$I_{\left(  4,8\right)  }^{A_{1}A_{2}A_{3}A_{4}}$ & $D_{A_{4}}^{A_{1}%
A_{2}A_{3}}$ & $\Delta$ & $N_{2-way}$ & $N_{3-way}$ & $N_{4-way}%
$\\\hline\hline
I & $\sum\limits_{K=2}^{4}\left(  \widehat{\rho}_{c}\right)  _{K}^{T_{A_{p}}%
}-2\widehat{\rho}_{c}$ & $\neq0$ & $\neq0$ & $\neq0$ & $\geq1$ & $\geq1$ &
$\geq1$\\\hline\hline
II & $\left(  \rho_{c}\right)  _{4}^{T_{A_{p}}}+\left(  \rho_{c}\right)
_{3}^{T_{A_{p}}}-\widehat{\rho}_{c}$ & $\neq0$ & $\neq0$ & $0$ & $0$ & $\geq1$
& $\geq1$\\\hline\hline
III & $\left(  \rho_{c}\right)  _{4}^{T_{A_{p}}}+\left(  \rho_{c}\right)
_{2}^{T_{A_{p}}}-\widehat{\rho}_{c}$ & $\neq0$ & $0$ & $\neq0$ & $\geq1$ & $0$
& $\geq1$\\\hline\hline
IV & $\left(  \rho_{c}\right)  _{4}^{T_{A_{p}}}$ & $\neq0$ & $0$ & $0$ & $0$ &
$0$ & $1$\\\hline\hline
V & $\left(  \rho_{c}\right)  _{3}^{T_{A_{p}}}+\left(  \rho_{c}\right)
_{2}^{T_{A_{p}}}-\widehat{\rho}_{c}$ & $0$ & $\neq0$ & $0$ & $\geq1$ & $\geq1$
& $0$\\\hline\hline
VI & $\left(  \rho_{c}\right)  _{3}^{T_{A_{p}}}$ & $0$ & $\neq0$ & $0$ & $0$ &
$1$ & $0$\\\hline\hline
VII & $\left(  \rho_{c}\right)  _{2}^{T_{A_{p}}}$ & $0$ & $0$ & $0$ & $\geq1$
& $0$ & $0$\\\hline\hline
\end{tabular}
\label{t1}%
\end{table}

A four qubit state with a single four way negativity font
\[
\left\vert \Psi_{ab}\right\rangle =a\left(  \left\vert 0000\right\rangle
+\left\vert 1111\right\rangle \right)  +b\left(  \left\vert 1101\right\rangle
+\left\vert 1110\right\rangle +\left\vert 0011\right\rangle \right)  \text{,}%
\]
is an example of class I states . Three qubit invariants for the state are
$I_{\left(  A_{4}\right)  _{0}}^{A_{1}A_{2}A_{3}}=a^{2}b^{2}$, $P_{\left(
A_{4}\right)  _{0}}^{A_{1}A_{2}A_{3}}=\frac{1}{2}a^{3}b$, $I_{\left(
A_{4}\right)  _{1}}^{A_{1}A_{2}A_{3}}=b^{4}$, $P_{\left(  A_{4}\right)  _{1}%
}^{A_{1}A_{2}A_{3}}=-\frac{1}{2}a^{2}b^{2}$, and $\left(  T^{A_{1}A_{2}A_{3}%
}\right)  _{A_{4}}=\frac{1}{6}\left(  a^{4}-2ab^{3}\right)  $. Four qubit
invariants are found to be $I_{(4,8)}^{A_{1}A_{2}A_{3}A_{4}}=\frac{1}%
{12}\left(  a^{4}+4ab^{3}\right)  ^{2}$, $D_{A_{4}}^{A_{1}A_{2}A_{3}}\neq0$,
and $\Delta\neq0$. A representative of class II states with $\widehat{\rho
}_{G}^{T_{A_{p}}}=\widehat{\rho}_{4}^{T_{A_{p}}}+\widehat{\rho}_{3}^{T_{A_{p}%
}}-\widehat{\rho}$ is, $\left\vert \Psi_{a}\right\rangle =a\left(  \left\vert
0000\right\rangle +\left\vert 1111\right\rangle \right)  +\left\vert
1110\right\rangle $. The state is SLOCC equivalent to GHZ state, however it
deserves a distinct status since on removal of qubit $A_{4}$ it has residual
three way coherences.

Invariants for class III states%
\begin{align}
G_{abcd}  &  =\frac{a+d}{2}\left(  \left\vert 0000\right\rangle +\left\vert
1111\right\rangle \right)  +\frac{a-d}{2}\left(  \left\vert 1100\right\rangle
+\left\vert 0011\right\rangle \right) \nonumber\\
&  +\frac{b+c}{2}\left(  \left\vert 1010\right\rangle +\left\vert
0101\right\rangle \right)  +\frac{b-c}{2}\left(  \left\vert 0110\right\rangle
+\left\vert 1001\right\rangle \right)  ,
\end{align}%
\begin{align}
L_{abc_{2}}  &  =\frac{a+b}{2}\left(  \left\vert 0000\right\rangle +\left\vert
1111\right\rangle \right)  +\frac{a-b}{2}\left(  \left\vert 1100\right\rangle
+\left\vert 0011\right\rangle \right) \nonumber\\
&  +c\left(  \left\vert 1010\right\rangle +\left\vert 0101\right\rangle
\right)  +\left\vert 0110\right\rangle ,
\end{align}

\begin{equation}
L_{a_{2}b_{2}}=a\left(  \left\vert 0000\right\rangle +\left\vert
1111\right\rangle \right)  +b\left(  \left\vert 0101\right\rangle +\left\vert
1010\right\rangle \right)  +\left(  \left\vert 0110\right\rangle +\left\vert
0011\right\rangle \right)  ,
\end{equation}
and%
\begin{equation}
L_{a_{2}0_{3\oplus\widetilde{1}}}=a\left(  \left\vert 0000\right\rangle
+\left\vert 1111\right\rangle \right)  +\left(  \left\vert 0101\right\rangle
+\left\vert 0110\right\rangle +\left\vert 0011\right\rangle \right)  ,
\end{equation}
of ref. \cite{vers02} with $\left(  \rho_{c}\right)  _{G}^{T_{A_{p}}}=\left(
\rho_{c}\right)  _{4}^{T_{A_{1}}}+\left(  \rho_{c}\right)  _{2}^{T_{A_{1}}%
}-\widehat{\rho}_{c}$ are listed in Table \ref{t2}. All three way coherences
are convertible to two way coherences as such three-way negativity fonts have
zero determinants. Four qubit entanglement occurs due to four-way and two-way
coherences. For all these states, the invariants $P_{\left(  A_{4}\right)
_{0}}^{A_{1}A_{2}A_{3}}$ and $P_{\left(  A_{4}\right)  _{1}}^{A_{1}A_{2}A_{3}%
}$ are identically zero. In Table \ref{t2}, for states in family $G_{abcd}$,
three qubit invariants used for the set $A_{1}A_{2}A_{3}$ are%
\[
T_{A_{4}}^{A_{1}A_{2}A_{3}}=\frac{1}{6}\left(  A-2B\right)  ,\left(
I_{3}^{A_{1}A_{2}A_{3}}\right)  _{\left(  A_{4}\right)  _{0}}=\left(
I_{3}^{A_{1}A_{2}A_{3}}\right)  _{\left(  A_{4}\right)  _{1}}=B,
\]
where%
\begin{equation}
A=\left(  a^{2}-b^{2}\right)  \left(  d^{2}-c^{2}\right)  ,B=\frac{1}%
{4}\left(  a^{2}-d^{2}\right)  \left(  b^{2}-c^{2}\right)  . \label{AB}%
\end{equation}
For states $G_{ab00}$ and $G_{00cd}$, $\Delta=0$. For states $L_{abc_{2}}$,
with $\left(  T^{A_{1}A_{2}A_{3}}\right)  _{A_{4}}=\frac{1}{6}\left(
a^{2}-c^{2}\right)  \left(  b^{2}-c^{2}\right)  $, $\left(  I_{3}^{A_{1}%
A_{2}A_{3}}\right)  _{\left(  A_{4}\right)  _{0}}=c\left(  a^{2}-b^{2}\right)
$, the value $\left(  I_{3}^{A_{1}A_{2}A_{3}}\right)  _{\left(  A_{4}\right)
_{1}}=0$ results in $\Delta=0$. A comparison of states $L_{abc_{2}}$ with
$a=c$ and L$_{ab_{3}}$ shows that the states are not SLOCC\ equivalent
\cite{li09} because the number of negativity fonts is not equal. However,
since four qubit correlations are null $\left(  I_{(4,8)}^{A_{1}A_{2}%
A_{3}A_{4}}=0\right)  $ for $L_{abc_{2}}$ with $a=c$ as well as L$_{ab_{3}}$,
these are subclasses of the same major class in correlation type based
classification, partially supporting the result of \cite{chte07}.

The families of states $L_{ab_{3}}$ and $L_{a_{4}}$ of ref. \cite{vers02} have
a similar global partial transpose composition. The value of degree
two\ invariant is $I_{4}=\frac{3a^{2}+b^{2}}{2}$ for $L_{ab_{3}}$ and
$I_{4}=2a^{2}$ for $L_{a_{4}}$ indicating that four-way coherences are
present. However, for the set of qubits $A_{1}A_{2}A_{3}$, only non zero three
qubit invariant is $\left(  I_{3}^{A_{1}A_{2}A_{3}}\right)  _{\left(
A_{4}\right)  _{1}}=\frac{a^{2}-b^{2}}{2}$ for L$_{ab_{3}}$ and $\left(
I_{3}^{A_{1}A_{2}A_{3}}\right)  _{\left(  A_{4}\right)  _{1}}=-4a^{2}$\ for
$L_{a_{4}}$. A finite $I_{4}$ but zero $I_{(4,8)}^{A_{1}A_{2}A_{3}A_{4}}$
indicates that the superposition contains a product of two qubit entangled
states. Four partite entanglement may, in this case, be detected by products
$\left(  N_{A_{4}}^{A_{1}A_{2}A_{3}}\right)  \left(  N_{A_{3}}^{A_{1}%
A_{2}A_{4}}\right)  $.

The states in families $L_{a_{2}b_{2}}$ and $L_{a_{2}0_{3\oplus\widetilde{1}}%
}$ \cite{vers02} have $\left(  N_{A_{4}}^{A_{1}A_{2}A_{3}}\right)
^{2}=2\left\vert I_{(4,8)}^{A_{1}A_{2}A_{3}A_{4}}\right\vert $. The states in
$L_{a_{2}b_{2}}$ and $L_{a_{2}0_{3\oplus\widetilde{1}}}$ differ from each
other in the number of two way negativity fonts with non-zero determinants.
Only non zero three tangle for the states $G_{abba}$ is $T_{A_{4}}^{A_{1}%
A_{2}A_{3}}$.\begin{table}[ptb]
\caption{Invariants for class III states $G_{abcd}$, $L_{abc_{2}}$,
$L_{a_{2}b_{2}}$ and $L_{a_{2}0_{3\oplus\widetilde{1}}}$ with $\left(
\rho_{c}\right)  _{G}^{T_{A_{p}}}=\left(  \rho_{c}\right)  _{4}^{T_{A_{1}}%
}+\left(  \rho_{c}\right)  _{2}^{T_{A_{1}}}-\widehat{\rho}_{c}$. A and B in
column II are as defined in Eq. (\ref{AB}).}%
\label{t2}
\begin{tabular}
[c]{||c||c||c||c||c||}\hline\hline
Invariant$\backslash$Class & $G_{abcd}$ & $L_{abc_{2}}$ & $L_{a_{2}b_{2}}$ &
$L_{a_{2}0_{3\oplus\widetilde{1}}}$\\\hline\hline
$\left(  N_{A_{4}}^{A_{1}A_{2}A_{3}}\right)  ^{2}$ & $\frac{1}{6}\left\vert
A-2B\right\vert ^{2}+2\left\vert B\right\vert ^{2}$ & $%
\begin{array}
[c]{c}%
\frac{1}{6}\left\vert \left(  a^{2}-c^{2}\right)  \left(  b^{2}-c^{2}\right)
\right\vert ^{2}\\
+\left\vert c\left(  a^{2}-b^{2}\right)  \right\vert ^{2}%
\end{array}
$ & $\frac{1}{6}\left\vert \left(  a^{2}-b^{2}\right)  ^{4}\right\vert $ &
$\frac{1}{6}\left\vert a^{8}\right\vert $\\\hline\hline
$I_{(4,8)}^{A_{1}A_{2}A_{3}A_{4}}$ & $\left(  \frac{1}{12}\left(  A-2B\right)
^{2}+B^{2}\right)  $ & $\frac{1}{12}\left(  a^{2}-c^{2}\right)  ^{2}\left(
b^{2}-c^{2}\right)  ^{2}$ & $\frac{1}{12}\left(  a^{2}-b^{2}\right)  ^{4}$ &
$\frac{1}{12}a^{8}$\\\hline\hline
$D_{A_{4}}^{A_{1}A_{2}A_{3}}$ & $\neq0$ & $\left\vert c\left(  a^{2}%
-b^{2}\right)  \right\vert ^{2}$ & $0$ & $0$\\\hline\hline
$\Delta$ & $\neq0$ & $0$ & $0$ & $0$\\\hline\hline
\end{tabular}
\end{table}The states $G_{a00a}$ and $G_{0bb0}$, with $\left(  \rho
_{c}\right)  _{G}^{T_{A_{p}}}=\left(  \rho_{c}\right)  _{4}^{T_{A_{p}}}$
belong to class IV in classification scheme based on correlation type. For
these states only non-zero three tangle is $T_{A_{4}}^{A_{1}A_{2}A_{3}}$,
therefore $I_{(4,8)}^{A_{1}A_{2}A_{3}A_{4}}\neq0$, $\Delta=0$ and $D_{A_{4}%
}^{A_{1}A_{2}A_{3}}=0$.

The global partial transpose has composition, $\left(  \rho_{c}\right)
_{G}^{T_{A_{1}}}=\left(  \rho_{c}\right)  _{3}^{T_{A_{1}}}+\left(  \rho
_{c}\right)  _{2}^{T_{A_{1}}}-\widehat{\rho}_{c}$, for class V states
$L_{0_{7\oplus\overline{1}}}$and $L_{0_{5\oplus\overline{3}}}$. In both cases
$I_{(4,8)}^{A_{1}A_{2}A_{3}A_{4}}=0$, while the product $\left(  N_{A_{4}%
}^{A_{1}A_{2}A_{3}}\right)  \left(  N_{A_{3}}^{A_{1}A_{2}A_{4}}\right)  \neq
0$. Two states differ in the the number of two-way negativity fonts with
non-zero determinants. Only non-zero invariant for Class VI state
$L_{0_{3\oplus\overline{1}}0_{3\oplus\overline{1}}}$ \cite{vers02} with
$\left(  \rho_{c}\right)  _{G}^{T_{A_{1}}}=\left(  \rho_{c}\right)
_{3}^{T_{A_{1}}}$ is $\left(  I_{3}^{A_{2}A_{3}A_{4}}\right)  _{\left(
A_{1}\right)  _{0}}$. The state has only three qubit entanglement. Class VII
with $\left(  \rho_{c}\right)  _{G}^{T_{A_{p}}}=\left(  \rho_{c}\right)
_{2}^{T_{A_{p}}}$ contains four qubit states with W-type entanglement
represented by $L_{a=0b_{3}=0}$ and separable states with entangled qubit
pairs, for example $G_{aaaa}$.

The polynomial invariant $I_{(4,8)}^{A_{1}A_{2}A_{3}A_{4}}$ is non-zero on
states $\left\vert \Psi_{ab}\right\rangle $, $G_{abcd}$, $L_{abc_{2}}$,
$L_{a_{2}b_{2}}$, $L_{a_{2}0_{3\oplus\widetilde{1}}}$, $G_{a00a}$ and
$G_{0bb0}$ and vanishes on states L$_{ab_{3}}$, $L_{a_{4}}$, $L_{0_{7\oplus
\overline{1}}}$, $L_{0_{5\oplus\overline{3}}},L_{0_{3\oplus\overline{1}%
}0_{3\oplus\overline{1}}}$ and $G_{aaaa}$. We define an entanglement monotone
to quantify four qubit correlations as%
\begin{equation}
\tau_{(4,8)}=4\left\vert \left(  12I_{(4,8)}^{A_{1}A_{2}A_{3}A_{4}}\right)
^{\frac{1}{2}}\right\vert , \label{tau48}%
\end{equation}
which is one on states with maximal entanglement due to four-body
correlations, finite on all states with entanglement due to four-body
correlations and zero otherwise. The subscript $(4,8)$ is carried on from
$I_{(4,8)}^{A_{1}A_{2}A_{3}A_{4}}$. One can verify that on four qubit GHZ
state%
\[
\left\vert GHZ\right\rangle =\frac{1}{\sqrt{2}}\left(  \left\vert
0000\right\rangle +\left\vert 1111\right\rangle \right)
\]
as well as cluster states \cite{brie01,raus01}
\[
\left\vert C_{1}\right\rangle =\frac{1}{2}\left(  \left\vert 0000\right\rangle
+\left\vert 1100\right\rangle +\left\vert 0011\right\rangle -\left\vert
1111\right\rangle \right)
\]%
\[
\left\vert C_{2}\right\rangle =\frac{1}{2}\left(  \left\vert 0000\right\rangle
+\left\vert 0110\right\rangle +\left\vert 1001\right\rangle -\left\vert
1111\right\rangle \right)  ,
\]%
\[
\left\vert C_{3}\right\rangle =\frac{1}{2}\left(  \left\vert 0000\right\rangle
+\left\vert 1010\right\rangle +\left\vert 0101\right\rangle -\left\vert
1111\right\rangle \right)  ,
\]
$\tau_{(4,8)}=1$ and $\left(  N_{A_{4}}^{A_{1}A_{2}A_{3}}\right)
^{2}=2\left\vert I_{(4,8)}^{A_{1}A_{2}A_{3}A_{4}}\right\vert $. So what is
different in cluster states? We recall the invariants $J^{A_{i}A_{j}}$ from
\cite{shar102}, the invariants that detect entanglement of a selected pair,
$A_{i}A_{j}$, of qubits in a four qubit state. For a GHZ state\ $J^{A_{i}%
A_{j}}=\frac{1}{4}$, for $\left(  i\neq j\right)  =1$ to $4$, while for a
cluster state all $J^{A_{i}A_{j}}$ \cite{shar102}, do not have the same value.
In canonical form, GHZ\ has a single four-way negativity font, while a cluster
state has two four-way negativity fonts besides also having two-way negativity
fonts (state reduction does not destroy all the coherences).

Another state proposed through a numerical search in ref. \cite{brow06} to be
a maximally entangled state is%
\[
\left\vert \Phi\right\rangle =\frac{1}{2}\left(  \left\vert 0000\right\rangle
+\left\vert 1101\right\rangle \right)  +\frac{1}{\sqrt{8}}\left(  \left\vert
1011\right\rangle +\left\vert 0011\right\rangle +\left\vert 0110\right\rangle
-\left\vert 1110\right\rangle \right)  ,
\]
However, on this state%
\[
T_{A_{4}}^{A_{1}A_{2}A_{3}}=\left(  I_{3}^{A_{1}A_{2}A_{3}}\right)  _{\left(
A_{4}\right)  _{0}}=\left(  I_{3}^{A_{1}A_{2}A_{3}}\right)  _{\left(
A_{4}\right)  _{1}}=\frac{1}{32},
\]%
\[
\left(  P_{3}^{A_{1}A_{2}A_{3}}\right)  _{\left(  A_{4}\right)  _{0}}=\left(
P_{3}^{A_{1}A_{2}A_{3}}\right)  _{\left(  A_{4}\right)  _{1}}=0,
\]
therefore $I_{(4,8)}^{A_{1}A_{2}A_{3}A_{4}}=\frac{1}{256}$, and $\tau
_{(4,8)}=\sqrt{\frac{3}{4}}$. On two excitation four qubit Dicke state%
\[
\left\vert \Psi_{D}\right\rangle =\frac{1}{\sqrt{6}}\left(  \left\vert
0011\right\rangle +\left\vert 1100\right\rangle +\left\vert 0101\right\rangle
+\left\vert 1010\right\rangle +\left\vert 1001\right\rangle +\left\vert
0110\right\rangle \right)  ,
\]
we have, $\tau_{(4,8)}=\frac{5}{9},$ while it is zero on four qubit W-state%
\[
\left\vert W\right\rangle =\frac{1}{2}\left(  \left\vert 0000\right\rangle
+\left\vert 1100\right\rangle +\left\vert 1010\right\rangle +\left\vert
1001\right\rangle \right)  .
\]
Four tangle $\tau_{4}$ also vanishes on $W-$like state of four qubits,
however, it fails to vanish on product of two qubit entangled states. Contrary
to $\tau_{(4,8)}$, a non zero $\tau_{4}$ does not ensure four-partite
entanglement. On four qubit state%
\begin{align}
\left\vert HS\right\rangle  &  =\frac{1}{\sqrt{6}}\left(  \left\vert
0011\right\rangle +\left\vert 1100\right\rangle +\exp\left(  \frac{i2\pi}%
{3}\right)  \left(  \left\vert 1010\right\rangle +\left\vert 0101\right\rangle
\right)  \right) \nonumber\\
&  +\frac{1}{\sqrt{6}}\exp\left(  \frac{i4\pi}{3}\right)  \left(  \left\vert
1001\right\rangle +0110\right)  ,
\end{align}
conjectured to have maximal entanglement in ref. \cite{higu00}, we have
$D_{\left(  A_{3}\right)  _{0}\left(  A_{4}\right)  _{1}}^{00}=D_{\left(
A_{3}\right)  _{1}\left(  A_{4}\right)  _{0}}^{00}=\frac{1}{6}$, and for
$4-$way negativity fonts $D^{0011}=\frac{1}{6}$, $D^{0001}=\frac{1}{12}\left(
1-i\sqrt{3}\right)  ,$ and $D^{0010}=\frac{1}{12}\left(  1+i\sqrt{3}\right)
$). Therefore%
\[
T_{A_{4}}^{A_{1}A_{2}A_{3}}=\left(  I_{3}^{A_{1}A_{2}A_{3}}\right)  _{\left(
A_{4}\right)  _{0}}=\left(  I_{3}^{A_{1}A_{2}A_{3}}\right)  _{\left(
A_{4}\right)  _{1}}=0,
\]%
\[
\left(  P_{3}^{A_{1}A_{2}A_{3}}\right)  _{\left(  A_{4}\right)  _{0}}=\left(
P_{3}^{A_{1}A_{2}A_{3}}\right)  _{\left(  A_{4}\right)  _{1}}=0,
\]
leading to $\tau_{(4,8)}=0$. However, the invariant $I_{(2,6)}^{A_{1}%
A_{2}A_{3}A_{4}}=1$ (Eq. (\ref{i26})) on $\left\vert HS\right\rangle $ and
takes value $\frac{27}{64}$ on four qubit $\left\vert W\right\rangle $ state.
It reflects the fact that a measurement on the state of a qubit, in
$\left\vert HS\right\rangle $ always leaves the three remaining qubits in a
three qubit W-state, whereas a similar measurement on a $\left\vert
W\right\rangle $ state yields a mixture of three qubit W-state with three
qubits in a separable state.

The choice $I_{(4,8)}^{A_{1}A_{2}A_{3}A_{4}}$ to quantify four qubit
correlations is also supported by the conclusions of \ \cite{endr06}, where
for \ a selected set of four qubit states, generator S of ref. \cite{luqu03}
has been shown to have the same parameter dependence as optimized Bell type
inequalities and a combination of global negativity and 2-qubit concurrences.

To summarize, degree 8, 12 and 24 four qubit invariants, expressed in terms of
three qubit invariants, have been obtained. One can continue the process to
higher number of qubits. Commonly, multivariate forms in terms of state
coefficients $a_{i_{1}i_{2}...i_{N}}$ are used to obtain polynomial invariants
for qubit systems. Our strategy is to write multivariate forms with relevant
$K-$qubit invariants as coefficients. The advantage of our technique is that
relevant invariants in a larger Hilbert space are easily related to invariants
in sub spaces as such to the structure of the quantum state at hand.
Construction of polynomial invariants for states other than the most general
state is a great help in classification of states. Our method can be easily
applied to determine the invariants for any given state. Entanglement monotone
that quantifies four qubit correlations can be used to quantify correlations
in pure and mixed (via convex roof extension) four qubit states.

This work is financially supported by CNPq Brazil and FAEP UEL Brazil.

\end{document}